\documentclass[10pt, journal, comsoc]{IEEEtran}

\usepackage{graphicx}
\graphicspath{ {./img/} }
\usepackage{cite}
\usepackage{array}
\usepackage{ifpdf}
\ifpdf
\else
\fi
\usepackage{amsmath}
\usepackage{amssymb}
\usepackage[cmintegrals]{newtxmath}
\usepackage{enumitem}
\usepackage{booktabs}
\usepackage[normalem]{ulem}
\useunder{\uline}{\ul}{}
\usepackage{lscape}
\vspace{\baselineskip}
\usepackage{tabulary}
\usepackage{tabularx}
\usepackage{colortbl}
\usepackage{longtable}
\usepackage{multirow}
\usepackage{xcolor}
\usepackage[modulo]{lineno}
\usepackage{amssymb}
\usepackage{placeins}
\usepackage{float}
\usepackage{scalefnt}
\usepackage{multicol, blindtext, graphicx}
\useunder{\uline}{\ul}{}
\usepackage[utf8]{inputenc}
\usepackage{pifont}
\newcommand{\cmark}{\ding{51}}%
\newcommand{\xmark}{\ding{55}}%
\usepackage{lipsum}

\begin{document}


\title{OPlaceRAN - a Placement Orchestrator for Virtualized Next-Generation of Radio Access Network}

\author{Fernando Zanferrari Morais, Gustavo Zanatta Bruno, Julio Renner,  Gabriel Almeida, \\ Luis M. Contreras, Rodrigo da Rosa Righi,  Kleber Vieira Cardoso, and Cristiano Bonato Both
\IEEEcompsocitemizethanks{
\IEEEcompsocthanksitem Fernando Zanferrari Morais, Gustavo Zanatta Bruno, Julio Renner, Rodrigo da Rosa Righi, and Cristiano Bonato Both are with the University of Vale do Rio dos Sinos (UNISINOS). E-mail: \{fmorais, zanattabruno, juliorenner\}@edu.unisinos.br, \{rrrighi, cbboth\}@unisinos.br
\IEEEcompsocthanksitem Gabriel Matheus and  Kleber Vieira Cardoso are with the Federal University of Goiás (UFG). E-mail:  \{gabrielmatheus,kleber\}@inf.ufg.br
\IEEEcompsocthanksitem Luis M. Contreras is with the Transport \& IP Networks - Systems and Network Global Direction, Telefónica CTIO Unit. E-mail: luismiguel.contrerasmurillo@telefonica.com
}
}


\maketitle

\IEEEpeerreviewmaketitle
\begin{abstract}
The fifth-generation mobile evolution enables transformations on Next-Generation Radio Access Networks (NG-RAN). The RAN protocol stack is split into eight disaggregated options combined in three network units, i.e., Central, Distributed, and Radio. Besides that, further advances allow the RAN functions to be virtualized on top of general-purpose hardware, using the concept of virtualized RAN (vRAN). The Combination of NG-RAN and vRAN results in vNG-RAN, which enables the management of the disaggregated units and protocols as a set of radio functions. However, the orchestration-based placement of these radio functions is a challenging issue since the best decision can be determined by multiple constraints involving RAN disaggregation, crosshaul networks requirements, availability of computational resources, etc. This article proposes OPlaceRAN, a vNG-RAN deployment orchestrator framed within the NFV reference architecture and aligned with the Open RAN initiative. OPlaceRAN supports the dynamic placement of radio functions focusing on vNG-RAN planning and is designed to be agnostic to the placement optimization model. To validate OPlaceRAN, we developed a prototype based on up-to-date cloud-native tools to deploy RAN using containerized virtualization using the OpenAirInterface emulator and considering two distinct functional splits (options 2 and 6). The evaluation is tested as proofs-of-concept in a real computing infrastructure using two different placement solutions. Our results reveal that OPlaceRAN is an effective cloud-native solution for containerized network functions placement and agnostic to the optimization model. Additionally, OPlaceRAN is up-to-date with the most advanced vNG-RAN design and development approaches, contributing to the evolution of the fifth-generation of mobile networks.

\end{abstract} 
\begin{IEEEkeywords}
NG-RAN, RAN disaggregation, NFV, CNF placement, crosshaul networks
\end{IEEEkeywords}
\section{Introduction}
\label{sec:intro}

The evolution of the fifth-generation of mobile networks is guided by transforming technologies and architectures into centric-software concepts to meet the new service demands, e.g., ultra-low latency and high data bit rate applications. For mobile operators, the softwarization process brings significant advantages, especially concerning time-to-market, usage of commercial off-the-shelf (COTS) hardware, total cost of ownership (TCO) savings, adoption of open-source initiatives, continuous development and operations (DevOps), and network automation \cite{zhang2018network,gsmamigration}. 
%
Based on software concepts, the Radio Access Network (RAN) is in transition towards a general adoption of virtualization (i.e., vRAN) from the current situation in which RAN infrastructures are basically monolithic and highly distributed \cite{liyanage2015software}. The most promising movement is led by the industry's Open RAN (O-RAN)\footnote{https://www.o-ran.org/} initiative, focusing on open and interoperable solutions \cite{o-ran1}. Further advances are aligned with the proposed transformations towards the Next-Generation RAN (NG-RAN) architecture which allows the functional split of the radio protocol stack into eight potential options. These options can be combined into up to three network elements: \textit{(i)} Central Unit (CU), \textit{(ii)} Distributed Unit (DU), and \textit{(iii)} Radio Unit (RU) \cite{gstr2018transport,marsch20185g,ngran2021etsi}. These split options are intended to enable radio features and improve cost efficiency in comparison with previous generations of mobile networks \cite{agrawal2017cloud,marsch20185g}.

The functional splitting of NG-RAN combined with the virtualization in vRAN enables the vNG-RAN architecture to provide flexibility at the time of deploying mobile access networks \cite{3gpp2017study,garcia2021ran}. 
This flexibility allows the dynamic placement of radio functions following a fine-grained network management approach \cite{sehier2019transport}. However, vNG-RAN introduces an unprecedented challenging problem since crosshaul transport networks (composed of backhaul, midhaul, and fronthaul networks) and computational resources (hosting the virtualized radio functions) have resource limitations, different resource dimensioning, and non-uniform resources consumption patterns \cite{gstr2018transport}. In this context, the Network Function Virtualization (NFV) architecture leads the development and standardization of Virtualized Network Functions (VNF) in mobile networks to cope with the deployment and dynamic placement of vNG-RAN. Moreover, Container Network Functions (CNF) can be applied for radio functions based on the cloud-native approach and vRAN characteristics \cite{o-ran1}. 

The NFV Orchestrator (NFVO) is an essential piece in the NFV reference model for placement and orchestration of the radio function chaining. NFVO is responsible for two main controls: Resource Orchestration (RO) and Network Service Orchestration (NSO), to support the VNF placement decision-making \cite{etsi1network,isg2012network}. Thus, the RO control concerns are related to the capacity and resources of the NFV Infrastructure (NFVI). Besides a relevant influence of the crosshaul networks (mainly latency and bit rate) capacity. Furthermore, the NSO manages the lifecycle of virtualized functions, particularly the chaining between them \cite{badulescu2019etsi}. In this context, the O-RAN initiative is inspired by the NFV architecture, including the proposed Service Management and Orchestration (SMO) framework to guide the decision-making of the radio VNF placement \cite{o-ran1}.



The state-of-the-art related to placement and orchestration for vNG-RAN is diverse. On the one hand, placement's concepts are widely evaluated and massively developed in exact and heuristic approaches \cite{murti2020optimization,murti:20,garcia2018fluidran,yusupov2018multi,molner2019optimization}. 
On the other hand, orchestration solutions are analyzed for generic functions not specific to the RAN environment. In this context, the article of Matoussi et al. \cite{matoussi20205g} is an exception in the literature. However, the authors investigated the predecessor architecture of NG-RAN (i.e., C-RAN) using the OpenAirInterface (OAI) tool.
In the industry, open-source orchestration platforms are being developed for generic VNFs \cite{mamushiane2019overview}, without specialization for vNG-RAN, to the best of our knowledge. The Open Network Automation Platform (ONAP) is an exception, working with O-RAN but without a concrete placement orchestration solution \cite{rodriguez2020cloud}. 
Moreover, it is fundamental to emphasize that a great tendency in the telecommunication industry standard applies the cloud-native approach \cite{compute2019etsi} with the virtualization-based on containers, adopting Kubernetes (K8S) as the main container orchestration tool \cite{CloudNativeFundation2020,k8sdocs}.


Based on the diverse presented in the literature and industry open-source projects, this article proposes the Orchestrator Placement RAN (\textbf{OPlaceRAN}), a vNG-RAN deployment orchestrator framed within the NFV reference architecture and aligned to the O-RAN SMO framework. OPlaceRAN supports the agnostic placement of radio functions, focusing on the problem of vNG-RAN planning. Moreover, OPlaceRAN is designed following the functional NFVO sub-blocks, considering the RO control named \textbf{RANPlacer}, a complementary optimization module named \textbf{RANOptimizer}, the NSO control called \textbf{RANDeployer} and, finally, a data repository referred as \textbf{RANCatalogs}. RANPlacer handles the whole orchestration process, including the processing of external input from the Network Operator (quantity of radio units), crosshaul topology capacity, NFVI  resources, 
and the alternative placement solutions stored in the RANCatalogs. RANOptimizer works with both exact and heuristics agnostic placement solutions aware of the functional split requirements. In this case, the agnostic solution is a strategy of vNG-RAN placement applied on the OPlaceRAN developed independently of the orchestrator. RANDeployer applies the virtualized radio functions addressed by the placement approaches according to the RANPlacer inputs according to the RAN CNFs also stored in the RANCatalogs. 
All the configuration, initialization, and validation processes of the virtualized radio functions are performed and activated by the RANDeployer.

The results obtained in our evaluation show that OPlaceRAN is a powerful orchestrator to leverage the placement of virtualized functions of vNG-RAN in NFV architectures. 
Our contributions can be summarized as follows:

\begin{itemize}
    
    \item The definition of a conceptual orchestrator aligned with the NFV architecture and consistent with the O-RAN SMO framework for agnostic placement solutions for vNG-RAN.
    
    \item The development of a real prototype aligned with the cloud-native approach to develop vNG-RAN's virtualized radio functions considering the three units of vNG-RAN, i.e., vCU, vDU, and vRU.
    
    
    \item The realization of a proof-of-concept with real computational NFVI 
    to provide the virtualized function placement analysis for vNG-RAN networks.
    
    
    \end{itemize}

The article is organized into the following sections. Initially, Section \ref{sec:back} introduces the concepts developed in the article. Section \ref{sec:archi} presents the OPlaceRAN architecture, and Section \ref{sec:proto} describes the prototype used in the experiment. Moreover, Section \ref{sec:eval} shows the developed prototype and the proposed evaluation environment. Finally, Section \ref{sec:rw} brings the related work and Section \ref{sec:conc} concludes the article.

\section{Virtualized NG-RAN Orchestration}
\label{sec:back}



The decomposition of the radio software stack into non-monolithic components aggregated in elementary functions is vital to improving RAN flexibility and processing performance for the 5G and beyond networks. The standardization initiatives such as 3GPP \cite{3gpp2017study} and IMT-2020/5G from ITU-T \cite{gstr2018transport} specify the number and types of functional splits. Therefore, the NG-RAN architecture aims to disaggregate and distribute radio functionality into three units (CU, DU, and RU). Moreover, the NG-RAN architecture flexibility allows some other configurations with less than three units, named DU and RU integration (DU and RU), C-RAN (CU and DU), and D-RAN (CU, DU, and RU in a monolithic way) \cite{gstr2018transport}. In a disaggregated NG-RAN, the crosshaul transport network provides communication paths among radio units composed of different segments such as backhaul (mobile core to CU), midhaul (CU to DU), and fronthaul (DU to RU). Each functional split depends on radio capacity (e.g., channel bandwidth, modulation, and others), imposing distinct requirements in terms of latency, and bit rate for the crosshaul network \cite{Larsen2019, Gavrilovska2020}. Consequently, the crosshaul must guarantee the latency and bit rate required according to the variety of radio functional splits \cite{sehier2019transport, Larsen2019}.

\begin{figure}[!ht]  
    \begin{center}
        \caption{NFV reference architecture.}
        \label{fig:nfv_architecture}
        \includegraphics[width=.5\textwidth]{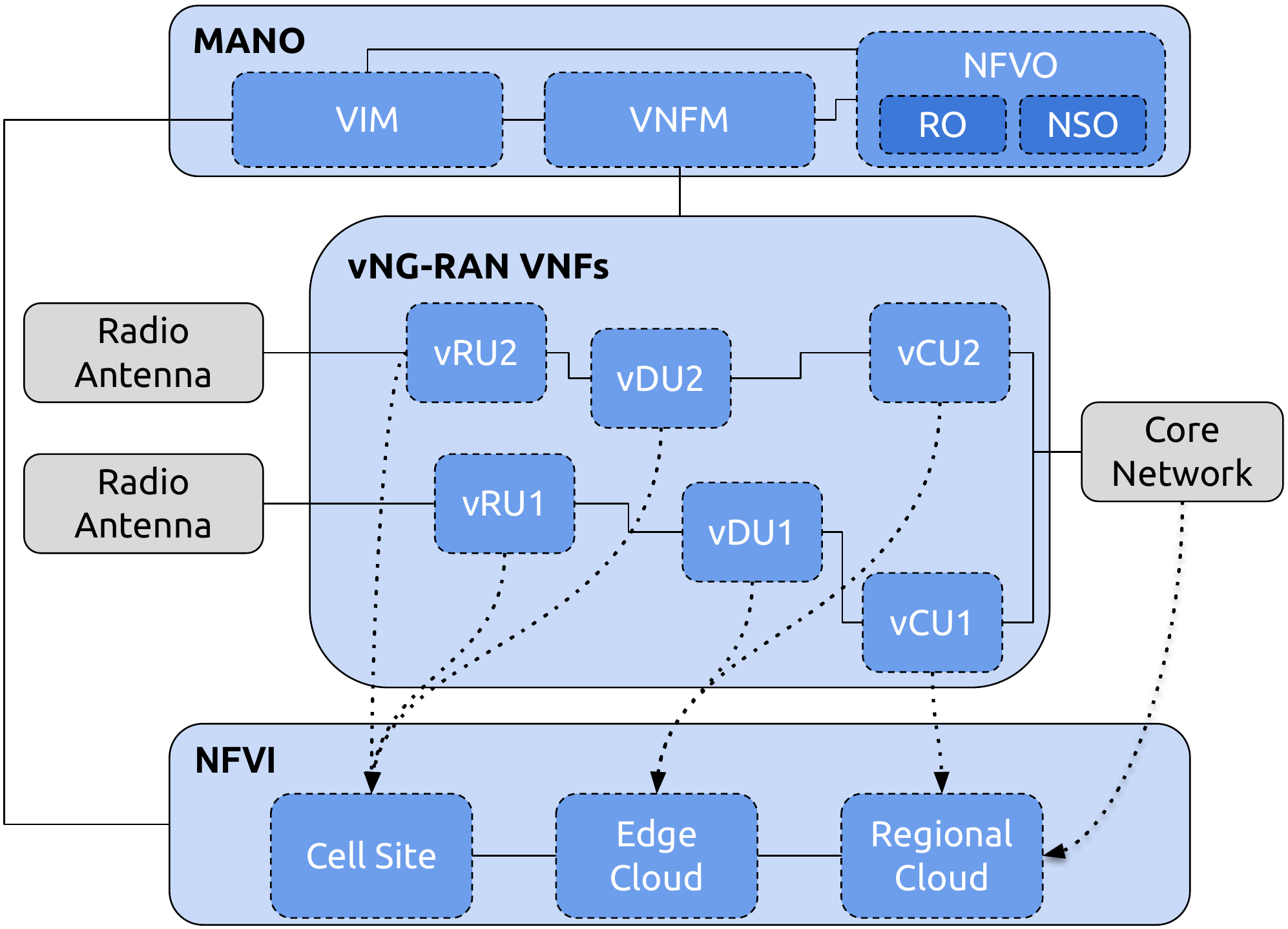}
    \end{center}
\end{figure}

The adoption of virtualization concepts is essential for RAN disaggregation. Each virtualized radio unit (vCU, vDU, and vRU) is a VNF, resulting in a vNG-RAN architecture. The main reference virtualization architecture for telecommunication is the NFV architecture specified by the European Telecommunications Standards Institute (ETSI) \cite{etsi1network}. In the NFV architecture, the Management and Orchestration (MANO) part is responsible for managing and orchestrating VNFs, defining such process as the automation, management, and operation of virtualized functions running on top of an infrastructure supporting virtualization and multi-tenancy. NFVO block orchestrates the NFVI resources across multiple Virtualized Infrastructures Managers (VIM) and manages the lifecycle of VNF services by the Virtual Network Functions Manager (VNFM). To this end, NFVO works with the RO control to provide access to the NFVI resources in an abstract manner, i.e., regardless of any VIM. Moreover, NFVO performs the NSO control to create end-to-end services, composing different VNFs, and managing the topology of network service instances \cite{bernardos2019network}. Fig. \ref{fig:nfv_architecture} presents the NFV architecture and the positioning of radio VNFs in the architecture. 

\begin{figure*}[!ht]  
    \begin{center}
        \caption{OPlaceRAN aligned with NFV architecture and set up with the O-RAN SMO framework.} 
        \label{fig:architecture}
        \includegraphics[width=.9\textwidth]{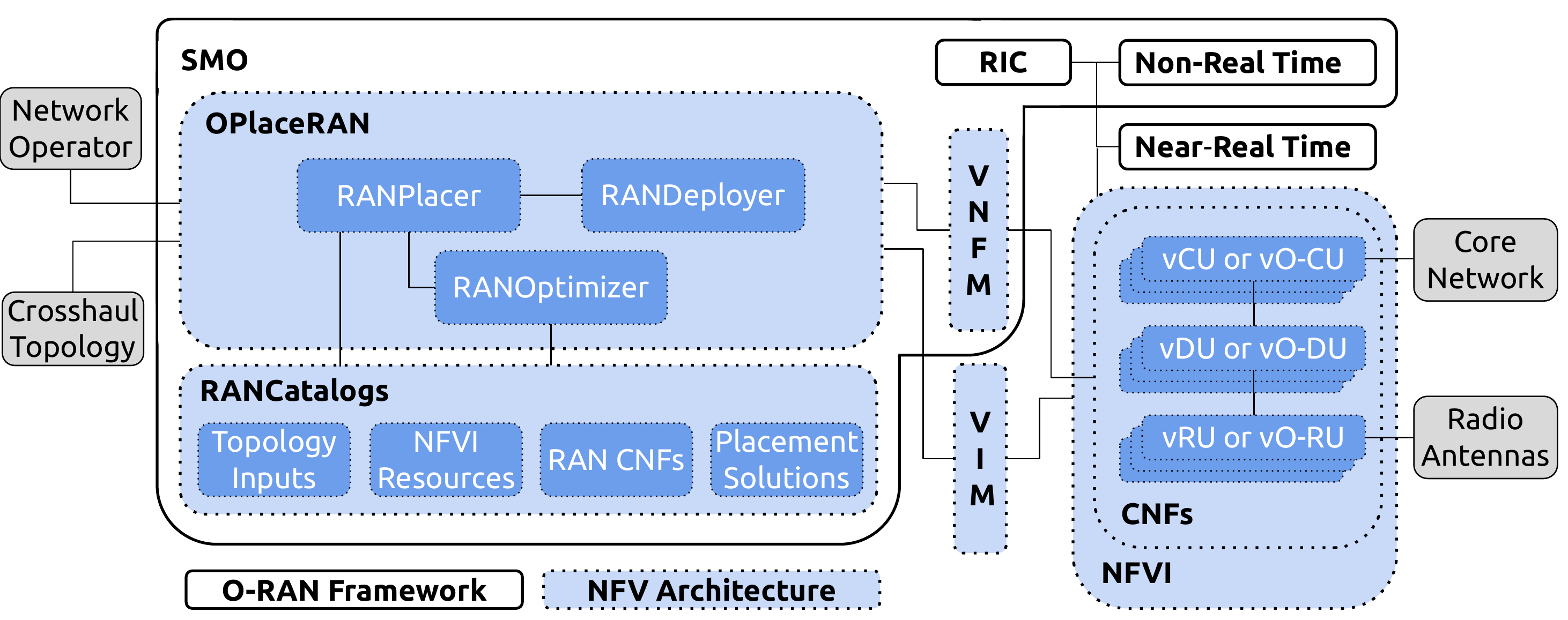}
    \end{center}
\end{figure*}

NFV architecture enables executing these radio VNFs over distributed Computational Resources (CR) interconnected through the crosshaul network. In this context, it is essential to highlight that O-RAN initiative also follows the NFV architecture. In this sense, O-RAN proposes the SMO framework focusing on slicing concepts, programmable RAN network resources, and embedded artificial intelligence capabilities \cite{o-ran1,garcia2021ran}. O-RAN uses both NFVO and VNFM MANO blocks to develop the orchestration and lifecycle of VNFs. In this sense, the SMO framework defines the RAN Intelligent Controller (RIC) to manage the RAN network divided into two components: (i) Non-Real Time (centralized components) and (ii) Near-Real Time (closed to vNG-RAN). Therefore, the SMO framework can connect to NFV architecture and provide a consolidated approach to the fifth-generation of mobile networks \cite{o-ran1,Gavrilovska2020}.   


To deploy RAN VNFs on top of NFVIs, the NFVO instantiate virtualized radio functions by leveraging hypervisor-based on virtual machines or containers. In this context, the container virtualization is targeted by O-RAN since it shares the Operating System that hosts the VNFs. 
Moreover, recent results show that the containerized solution tends to be the most appropriate for RAN virtualization due to the strict requirements of delay and scalability of resources. In this sense, each virtualized radio function is envisioned as a container \cite{rehman2019network, yi2018comprehensive,Gavrilovska2020,bonati2020open}. For the orchestration of containers in generic applications, four tools lead the implementations: \textit{(i)} K8S, \textit{(ii)} Docker Swarm, \textit{(iii)} Mesosphere Marathon, and \textit{(iv)} Cloudify. From the four options, K8S emerges in telecommunications virtualization as the main tool due to extensive Continuous Integration/Continuous Delivery (CI/CD) development, flexible configuration options (e.g., customized thresholds), and support for Docker-based containers (widespread). K8S focuses on managing virtualized infrastructures and orchestration in the form of clusters managed over a pool of computational resources, whether physical or virtual \cite{CloudNativeFundation2020}. According to that, the vNG-RAN placement orchestration solution is explored in the next section.




\section{OPlaceRAN: Orchestrator Placement RAN}
\label{sec:archi}


This section introduces OPlaceRAN, i.e., an orchestrator with decision-making support for radio function placement. We design OPlaceRAN to overcome the lack of placement orchestrators for vNG-RAN. OPlaceRAN  seeks to be an agnostic placement tool considering exact and heuristic placement solutions for the orchestration of planning of vNG-RAN. According to the NFVO functional block of the NFV architecture, OPlaceRAN operates between RO and NSO. In this context, OPlaceRAN is aware of crosshaul topologies. Moreover, the defined virtualization technology is based on containers, following O-RAN initiative to obtain the possibility of interoperability with tools that meet the evolution of the fifth-generation of mobile networks.


Fig. \ref{fig:architecture} shows OPlaceRAN positioned into NFV reference architecture for the RAN planning and design in the SMO framework. The overall NFV architecture components are covered, including VNFM, VIM, CNFs, and NFVI blocks in the end-to-end orchestration structure of vNG-RAN. In this sense, the OPlaceRAN design represents the intersection between the NFV architecture and the SMO framework, since SMO adopts NFVO and VNFM blocks from the NFV architecture. Moreover, it is worth to highlight that Near-Real Time is positioned outside the SMO framework, i.e., close to vNG-RAN (vCU/vO-CU, vDU/vODU, and vRU/vO-RU). Lastly, it connects to the Core Network and Radio Antennas (including UEs) and receives external input data from Network Operator and Crosshaul Topology.

OPlaceRAN is responsible for the entire orchestration of the system to place virtualized radio functions into NFVI, playing as an NFVO. The main goals of OPlaceRAN are: \textit{(i)} receive input from Network Topology by Network Operator, \textit{(ii)} analyze the available computational resources (CRs), \textit{(iii)} support an optimization solution for planning RAN networks, and \textit{(iv)} execute the deployment for positioning the virtualized radio functions. Moreover, OPlaceRAN interacts with the entire system. For example, OPlaceRAN communicates with VIM and VNFM blocks to collect infrastructure information and execute network scaling actions regarding MANO functions. In this context, OPlaceRAN adopts the indirect communication mode between VIM and VNFM blocks, as indicated in the NFV architecture \cite{etsi1network}.
For providing this indirect mode, OPlaceRAN is divided into three functional sub-blocks: RANPlacer, RANOptimizer, and RANDeployer. The three functional sub-blocks are described as following.


\begin{itemize}
    \item \textbf{RANPlacer}: it is positioned as a RO functionality from the NFVO perspective, centralizing and determining the actions of OPlaceRAN for making decisions. To begin the vNG-RAN planning process, RANPlacer operates on demand from external inputs. RANPlacer receives external vRUs from Network Operator and Crosshaul Topology. Based on these external inputs, RANPlacer requests VIM block to update NFVI Resource Catalog. These inputs, together with NFVI resources, are sent to the RANOptimizer sub-block to analyze the optimal placement of the virtualized radio functions. The optimized result is analyzed by RANPlacer, which sends the VNF allocation and chaining plans to the RANDeployer sub-block.

    
    \item \textbf{RANOptimizer}: it is responsible for the decision-making process for the placement of virtualized radio functions. In this case, RANOptimizer receives requests from, and returns solutions to, RANPlacer. This request is made by inserting the RANCatalogs sub-block inputs and choosing a placement solution for runtime deployments. After that, RANOptimizer delivers the result of the radio VNF placement solution. In this case, RANPlacer receives from RANOptimizer a solution when feasible, including \textit{(i)} position of vRU, vDU, and vCU units, \textit{(ii)} disaggregation RAN options, and \textit{(iii)} CRs to be allocated in the crosshaul networks.

    \item \textbf{RANDeployer}: it allocates the virtualized radio functions according to placement results from RANOptimizer sent to RANPlacer. In this case, RANDeployer operates as an NSO in NFVO, interacting with both VNFM and VIM blocks to deploy radio units. The interaction with the VNFM functional block assigns the creation and chaining activities of CNFs for virtualized radio functions. In addition, the interaction with VIM pursues the allocation of CNFs in the resources of CRs. Therefore, RANDeployer processes the requests from RANPlacer, requests radio units CNFs from the RANCatalogs sub-block, and sends the positions of CNFs to VNFM, deploying them in CRs of the NFVI.
    

\end{itemize}

\begin{figure*}[!ht]  
    \begin{center}
        \caption{OPlaceRAN diagram sequence.}
        \label{fig:sequence}
        \includegraphics[width=0.9\textwidth]{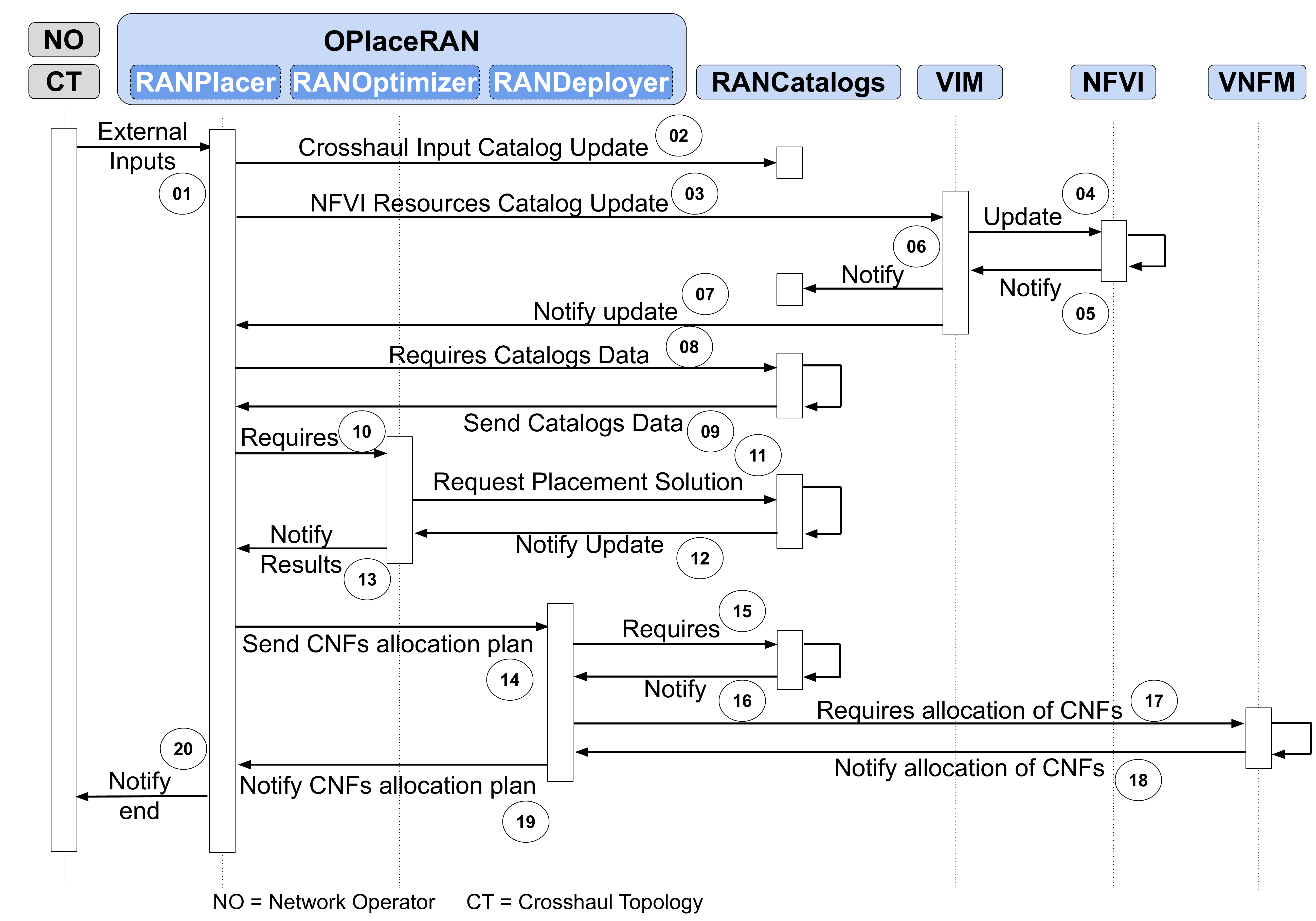}
    \end{center}
\end{figure*}

    To provide data support to OPlaceRAN operations, the \textbf{RANCatalogs} repository stores data about the overall services, functions, and resources available on the network. The RANCatalogs repository is composed of four different sub-catalogs. The first and second, i.e., Topology Inputs and NFVI Resources, operate as repositories of configurations. The Topology Inputs catalog works with external data entries from the Crosshaul Topology inputs, e.g., network links and their respective latency and data rate capacities. Therefore, the Topology Inputs catalog is updated by the RANPlacer. The NFVI resources catalog refers to the available and allocated computing resources in NFVI. Examples of data types found in NFVI resources are CPU, memory, and storage. In this case, the update is based on the communication between RANPlacer and VIM. The third Catalog sub-block is Placement solutions. This sub-block supports a pool of the agnostic exact and heuristic approaches of placement solutions requested by RANOptimizer. Lastly, the RAN CNFs catalog operates as a repository of the container image registry for the radio units, i.e., vCU, vDU, and vRUs. The request for CNFs is triggered by the RANDeployer sending them to VNFM aiming to access the radio units and allocate them according to the guidance of the RANDeployer allocation.

Finally, Fig. \ref{fig:sequence} presents the sequence diagram of the OPlaceRAN workflow. The process starts with the collection of External Inputs event (step 01) of the Network Operator (NO) and Crosshaul Topology. Next, RANPlacer processes those inputs by considering the vRUs entry and the Crosshaul Topology (step 02). After that, RANPLacer sends a request to RANCatalog updating the VIM CRs (step 03), asking for an up-to-date view to NFVI (step 04 and 05) and receiving the update notification (steps 06 and 07). In the following interaction, OPlaceRAN requires an update for the Topology Inputs and NFVI Resources to the RANCatalog (steps 08 and 09). Therefore, RANPlacer sends the updated information and the external data to RANOptimizer, selecting the placement solution of RANCatalog and providing a feasible radio function placement (steps 10, 11, 12, and 13). After that, RANPlacer addresses the allocation plan to the RANDeployer function (step 14). Lastly, RANDeployer requires RANs CNFs Catalog images and sends them to VNFM for the CNFs chaining allocation (steps 15, 16, 17, and 18). Furthermore, the sequence diagram shows the interaction of the completion of the tasks between the NFVO module and the rest of components of the whole NFV architecture through the VIM, NFVI, and VNFM. The OPlaceRAN operation finishes with the allocation of the VNFs in CRs and its notification (steps 19 and 20). The following section describes a prototype to validate the OPlaceRAN approach on K8S, considering this sequence diagram discussed.

\section{OPlaceRAN Prototype}
\label{sec:proto}

\begin{figure}[!ht]
    \begin{center}
        \caption{OPlaceRAN prototype.}
        \label{fig:prototype}
        \includegraphics[width=.45\textwidth]{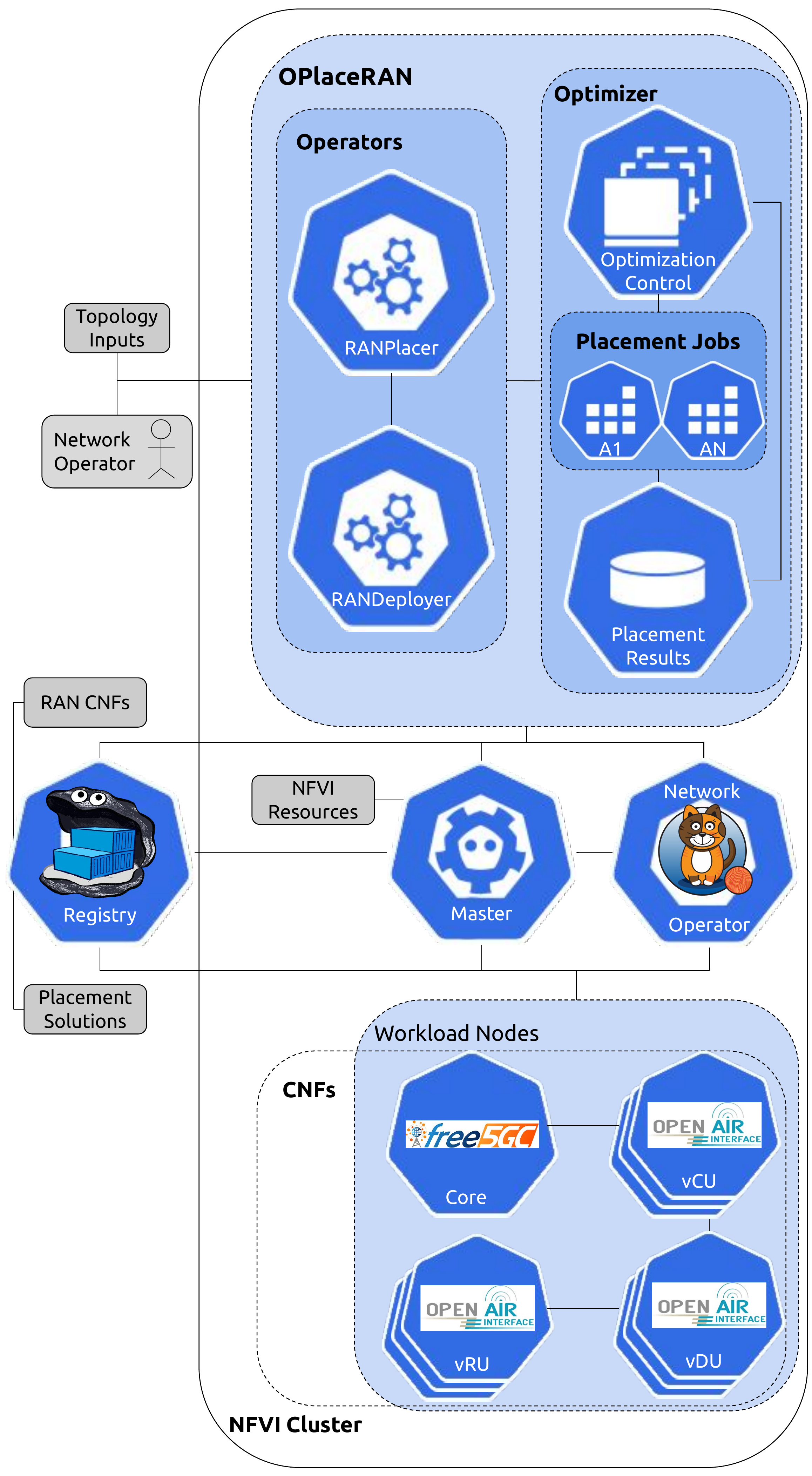}
    \end{center}
\end{figure}

To validate the OPlaceRAN proposition, we developed a K8S cloud-native tool prototype. Fig.  \ref{fig:prototype} shows the prototype components which comprise the NFV and SMO blocks to support OPlaceRAN and their sub-blocks RANPlacer, RANOptimizer, RANDeployer, and RANCatalog data repository. The prototype covers all the aforementioned presented workflow, from the collection of external inputs (NO inputs and Crosshaul Topology) up to the CNFs chaining allocation process. In the prototype, RANCatalog maintains the Topology Inputs and the NFVI resources implemented with custom resources feature from Kubernetes and uses a Container Registry for the CNF’s repository. The K8S tools include both NFVI and VIM natively and support VNFM and NFVO of the NFV architecture. Moreover, the K8S cluster comprises nodes concerning CRs and runs either directly on top of the hardware (bare-metal approach) or abstracted as a virtual machine (VM). In this context, applications are deployed and managed by K8S into Pod resources, holding one or more containers, i.e., CNFs. In the K8S architecture, Master Nodes provide the complete control of the cluster through a K8S control plane deployed in specific nodes. However, CNFs are running in the Workload Nodes. The main components of this control plane are Controller Manager, Scheduler, Storage, and APIs. The NFVI cluster (K8S oriented) applied in the prototype is represented by VMs. The Calico plugin is used as a Container Network Interface (CNI) to provide an IP network connection between the Pods (i.e., radio functions). The entire prototype code is available at GitHub\footnote{https://github.com/my5G/OPlaceRAN}.



K8S is highly customizable, even considering its default components. For example, the control plane works with the Operators pattern to extend the Controller Manager and APIs. The operator pattern combines two extension points: the API server extension adding new resources to the K8S API and a controller monitoring and executing the defined resources. Therefore, an Operator can automate the whole lifecycle of the prototype, providing application-specific automation without changing the K8S default behavior. Moreover, Operators work under a Pod and are driven by the K8S object model. K8S uses additional layers of abstraction over the container interface to provide scale, resiliency, and lifecycle management capabilities. In this sense, both RANPlacer and RANDeployer orchestration sub-blocks are developed under the K8S operator’s pattern in the Golang language. Operators manage the decision-making process of the solution. For example, RANPlacer builds the information base (i.e., the VIM data and placement result) for the RANDeployer to allocate CNFs under Pods.

The Network Operator information includes the demand for vRUs and the Crosshaul Topology. The vRU demand is defined by the cluster of CR nodes, their vRU quantity, and the Crosshaul Input, one of three Catalog blocks defined as a ConfigMap. In the Catalog sub-block, NFVI Resources are acquired natively from K8S and matched with CRs of the nodes, i.e., the CPU and RAM capacity. The Placement Solution and the RAN CNFs Catalogs components are deployed in a Docker Registry server hosted in a virtual machine out of the NFVI cluster. The Docker Registry stores containerized images storing the placement solutions and the CNFs OAI images. These images are deployed on-demand by OPlaceRAN. We created images from available open-source projects: free5GC \cite{free5GC} for Core Network and OpenAirInterface (OAI) \cite{OAI} for the RAN units. We only disabled the PHY layer to emulate RAN, considering all other layers to support the disaggregation on three units (vCU, vDU, and vRU), using the split options 2 (O2 - F1 interface) and 6 (O6 - nFAPI interface).

The RANOptimizer works based on three sub-blocks to deliver the placement results to the RANDeployer: (i) Optimization Control, (ii) Placement Jobs, and (iii) Placement Results. The Optimization Control receives requests from the RANPlacer with the inputs and the Placement Solutions. The input is composed of the following pieces of information: (i) CR nodes, (ii) number of chains that must be deployed, (iii) placement of vRUs in CRs, (iv) crosshaul topology, (v) capacity of the CRs (CPU and RAM), and (vi) specific placement solution (either the exact or heuristic approaches). The Optimization Control works similarly to an HTTP server. Therefore, the Optimization Control asynchronously triggers the placement solution job execution and provides to the RANPlacer a token used to get the placement solution status and the result. The Placement Jobs are also based on asynchronous tasks since they can take a considerable time to finish. In this case, a Job resource handles the execution of the placement solution, starting one or more Pods to provide a task-based workflow expected to run successfully after completing their goal. According to the selected placement solution by the Network Operator, a job is triggered by the chosen solution. The Placement Solutions are deployed as container images stored in the Registry repository. Once the job finishes its execution, it saves the placement results to the persistency layer accessible by the Optimization Control. Finally, the result of the placement solution is returned to the RANPlacer to be suitable for sending towards RANDeployer for the CNFs allocation process. In the next section, we evaluate the prototype in an experimental topology.

\section{Evaluation}
\label{sec:eval}


This section evaluates OPlaceRAN in a Proof-of-Concept. Subsection \ref{subsec:proof-concept} provides a general description of the evaluated scenario, detailing the placement solution applied, including topology, and resources, by varying parameters to assess the solution results. Subsection \ref{subsec:results} presents the results obtained in the evaluation of OPlaceRAN concerning placement solution results, cluster, and nodes behavior. Lastly, Subsection \ref{discussion} discuss general aspects and relevant characteristics of OPlaceRAN. 


\subsection{Scenario Description}\label{subsec:proof-concept}

We organized the assessment of OPlaceRAN into four parts, as described in the following. The first part presents the topology used, while the second shows the computational capacity of the resources. In the third part, the parameters used in vNG-RAN are explored. Finally, the fourth part details the placement solutions used.


\begin{itemize}



\item \textbf{Topology}: we run the experiments in a K8S cluster in version 1.19.3, with six nodes. Each node of this cluster is hosted in a VM running on a VMware ESXi 6.7 hypervisor. The six VMs run Ubuntu 18.04 with a low latency kernel, in which two nodes have the master control plane (no vNG-RAN was deployed at this stage) and the other four nodes have the worker role. VMs are hosted in three DELL PowerEdge M610 servers equipped with two Intel Xeon X5660 processors running at 2.80 GHz and 192 GB of RAM. The Crosshaul Topology applied in the evaluation includes a physical switch (pSwitch) connected with the servers through a virtual switch (vSwitch) in the border. Each crosshaul link has a 10 Gbps reserved bandwidth and 1ms of latency guaranteed, with exception for the worker node six, which connects directly with the pSwitch and has 1 Gbps of bandwidth. Links latency between vSwitches and pSwitches are not considered. \ref{fig:topology} shows the topology used, which supports four NG-RAN scenarios considering the units and nodes distribution, as addressed in the RAN Units item (described below).

\begin{figure}[h!]
    \begin{center}
        \caption{Experimental topology.}
        \label{fig:topology}
        \includegraphics[width=.45\textwidth]{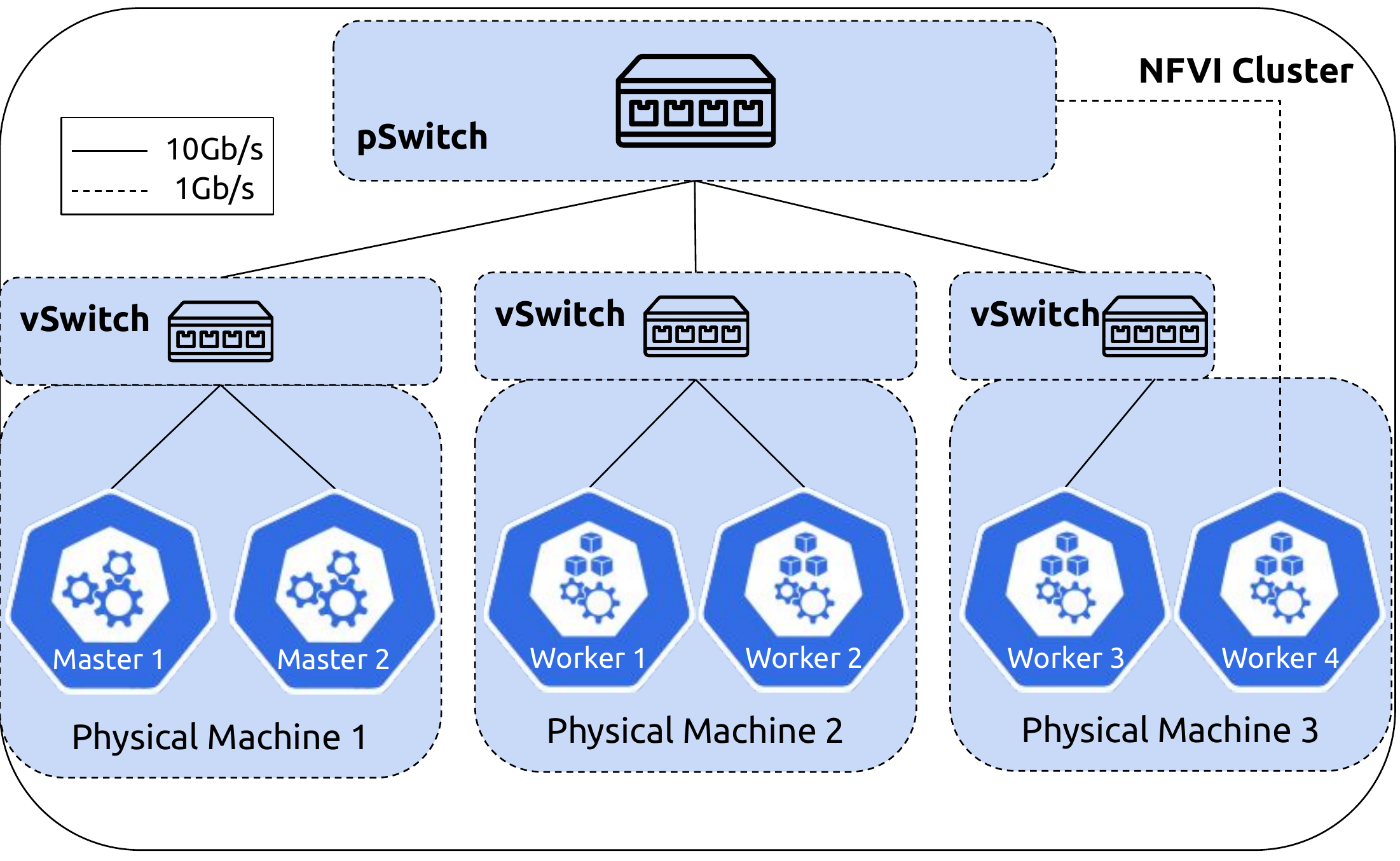}
    \end{center}
\end{figure}


\item \textbf{Computing Resources}: we focused on the processing capacity (i.e., CPU and RAM) because this metric is the most common bottleneck for computing devices in the context of vRAN evaluation, according to the OAI analysis \cite{lin2018performance,hu2019architectural}. CR values in Table \ref{tab:computing_resouces} indicate the number of available CPUs (Core unit) and memory RAM (GB unit) in worker K8S nodes at the time of generating input data. We fixed these values to have control under the deployment scenario, but our operator can automatically collect these values in the real world in placement time. The heterogeneity of CRs capacity used adds some complexity to the placement solutions that we evaluate. These capacities of CRs considered the CPU utilization profile of the RAN software as obtained from OAI implementation. The exact values may vary according to the adopted software components and the computing device, but similar profiles are employed in different works \cite{lin2018performance,hu2019architectural}.

\begin{table}[h!]
    \caption{Resources to RAN CNFs on K8S Work Nodes.}
    \centering
    \label{tab:computing_resouces}
        \begin{tabular}{|c|c|c|c|c|}
        \hline
        \textbf{Node}        & \textbf{1} & \textbf{2}   & \textbf{3}   & \textbf{4}   \\ \hline
        \rowcolor[HTML]{cdd5e4} CPU (Core number)        & 5 & 1   & 2   & 1 \\
        Memory(GB)  & 2 & 0.2 & 1.2 & 0.2 \\ \hline
    \end{tabular}
\end{table}


\item \textbf{RAN Units}: We defined a vRU for each work node. The parameters used in the two deployed placement models follow the definitions of the 3GPP \cite{3gpp2017study}, and ITU-T standardizing organizations \cite{gstr2018transport}, and the OAI Long Term Evolution (LTE) tool support, according to Table \ref{tab:profileslte}. We use four NG-RAN scenarios regarding the units and nodes distribution, i.e., \textit{(i)} three units divided into vCU, vDU, and vRU; \textit{(ii)} two units co-located vCU and vDU, \textit{(iii)} C-RAN scenario, using two units and integration of vDU and vRU; and \textit{(iv)} D-RAN with all the units in a single node. Additional parameters shown in Table \ref{tab:profileslte} are latency and bandwidth supported by the OAI emulator for RAN functional split. In this sense, the Core Network (CN) parameters represent the requirements of communication between CN and vCU. In this work, we apply two RAN split options, namely O2 and O6 \cite{small2016small}. In Table \ref{tab:profileslte}, the splits are represented by parameters O2 and O6. The O2 split parameters represent the vCU-vDU communication, and the O6 split parameters reflect the vDU-vRU communication.

\begin{table}[!h]
    \centering
    \caption{LTE parameters of RAN functional split \cite{small2016small}.}
    \label{tab:profileslte}
            \begin{tabular}{|c|c|c|c|c|c|c|}
            \hline
            \multirow{2}{*}{\textbf{NG-RAN Scenarios}} & \multicolumn{3}{c|}{\textbf{Latency - one way (ms)}} & \multicolumn{3}{c|}{\textbf{Bandwidth (Mbps)}} \\ \cline{2-7} 
              & \textbf{CN}      & \textbf{O2}     & \textbf{O6}     & \textbf{CN}    & \textbf{O2}   & \textbf{O6}   \\ \hline
            \rowcolor[HTML]{cdd5e4} NG-RAN 3 units & 30 & 30 & 2 & 151 & 151 & 152 \\
            NG-RAN 2 units & 30 & 30 & - & 151 & 151 & -   \\
            \rowcolor[HTML]{cdd5e4}  C-RAN          & 30 & -  & 2 & 151 & -   & 152 \\
            D-RAN          & 30 & -  & - & 151 & -   & -   \\ \hline
        \end{tabular}
\end{table}


\item \textbf{Placement Solutions}: we used two different placement solutions approaches to test the effectiveness of OPlaceRAN and the support for agnostic placement solutions. First PlaceRAN \cite{morais2021placeran}, an exact optimization approach focused on maximizing the aggregation level of RAN CNFs and minimizing the number of CRs used for such aggregation. The second is another exact optimization approach that combines a linearization technique with a cutting-planes method \cite{murti2020optimization}. The focus is to apply the Multi-CUs vRAN concept to minimize the vRAN cost and overall routing based on CU’s positioning. Both approaches explore the optimal solution of placement. It is fundamental to highlight that our goal is not to investigate the performance evaluation of placement solutions but the agnostic support for these solutions. 

\end{itemize}


\subsection{Results} \label{subsec:results}


We evaluate OPlaceRAN as proof-of-concept to obtain results that could indicate the scenario behavior using the orchestrator developed. Therefore, first, we detail the RAN CNFs allocation plan based on the placement solutions presented in the previous subsection. Second, we present the behavior of clusters and nodes. Finally, to evaluate the cluster’s performance, we use just the PlaceRAN \cite{morais2021placeran} placement solution because the focus is just the behavior of OPlaceRAN and not the performance of the placement solutions.

\begin{figure}[h!]
    \begin{center}
        \caption{Comparison of Multi-CU vRAN \cite{murti2020optimization} vs PlaceRAN \cite{morais2021placeran} solutions.}
        \label{fig:topology-results}
        \includegraphics[width=.5\textwidth]{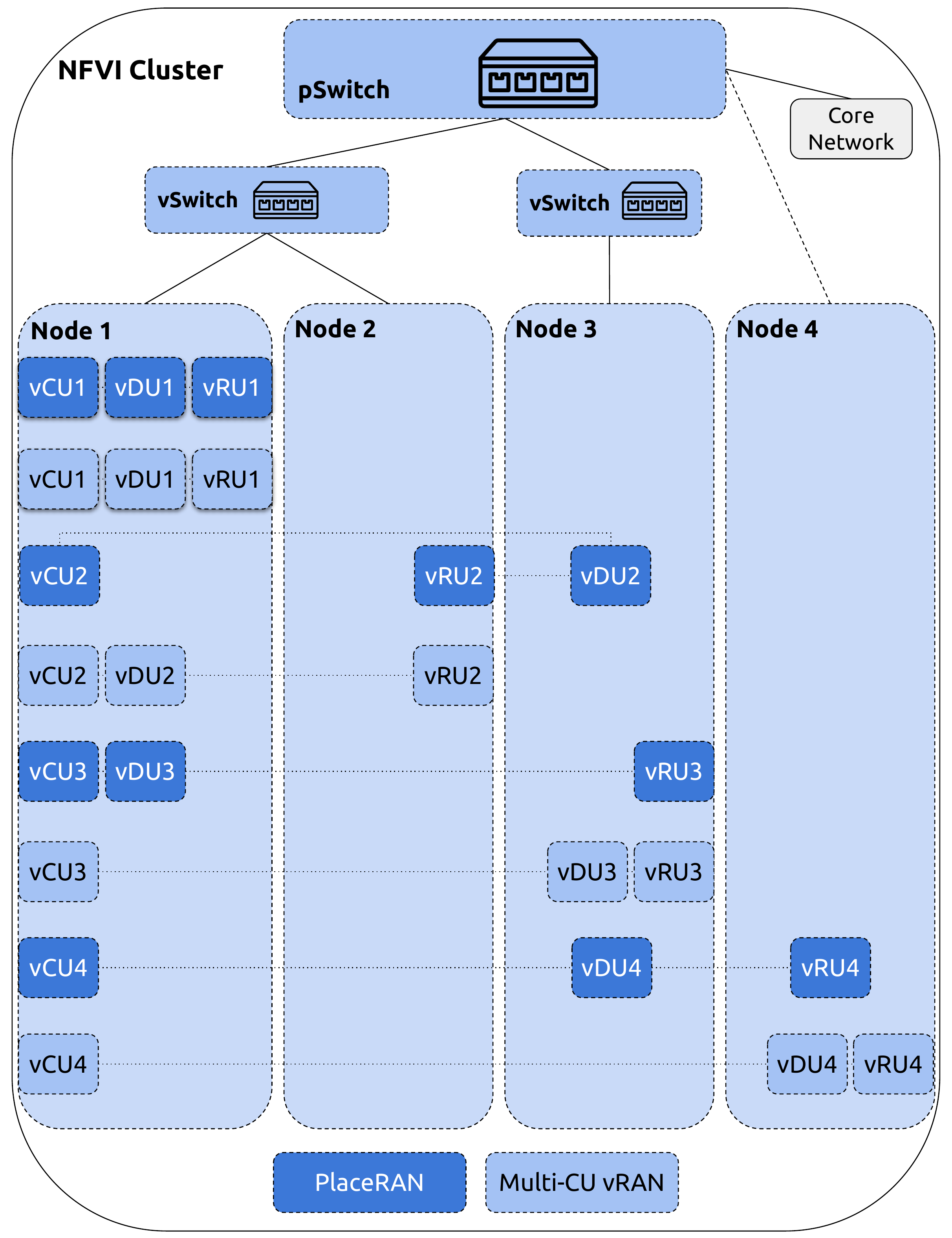}
    \end{center}
\end{figure}

Regarding the level of aggregation of CNFs, which indicates the capacity of putting RAN unit nearest to the Core Network and in the same CR, respecting the requirements of crosshaul networks, mainly driven by latency and bit rate \cite{morais2021placeran}. The results for the two placement solutions, PlaceRAN \cite{morais2021placeran}, and Multi-CUs vRAN \cite{murti2020optimization} are presented in Fig. \ref{fig:topology-results}. The mathematical modeling of these placement solutions uses the same number of nodes, and each node operates with only one RU (in the present work, only vRUs). PlaceRAN presents a slightly better performance because it aggregates more CNFs than the Multi-CUs vRAN model. Moreover, PlaceRAN concentrates on aggregating vCUs and vDUs CNFs only in Nodes 1 and 3, while Multi-CUs vRAN aggregates in Nodes 1, 3, and 4. In this analysis, PlaceRAN presents a better performance in the centralization of CNFs than Multi-CUs vRAN. This centralization is due to the flexible allocation of vDUs designed in PlaceRAN. The placement of DUs is predefined and inflexible in multi-CUs vRAN.



\begin{figure}[h!]
    \begin{center}
        \caption{Initial OPlaceRAN steps.}
        \label{fig:operator-times}
        \includegraphics[width=.5\textwidth]{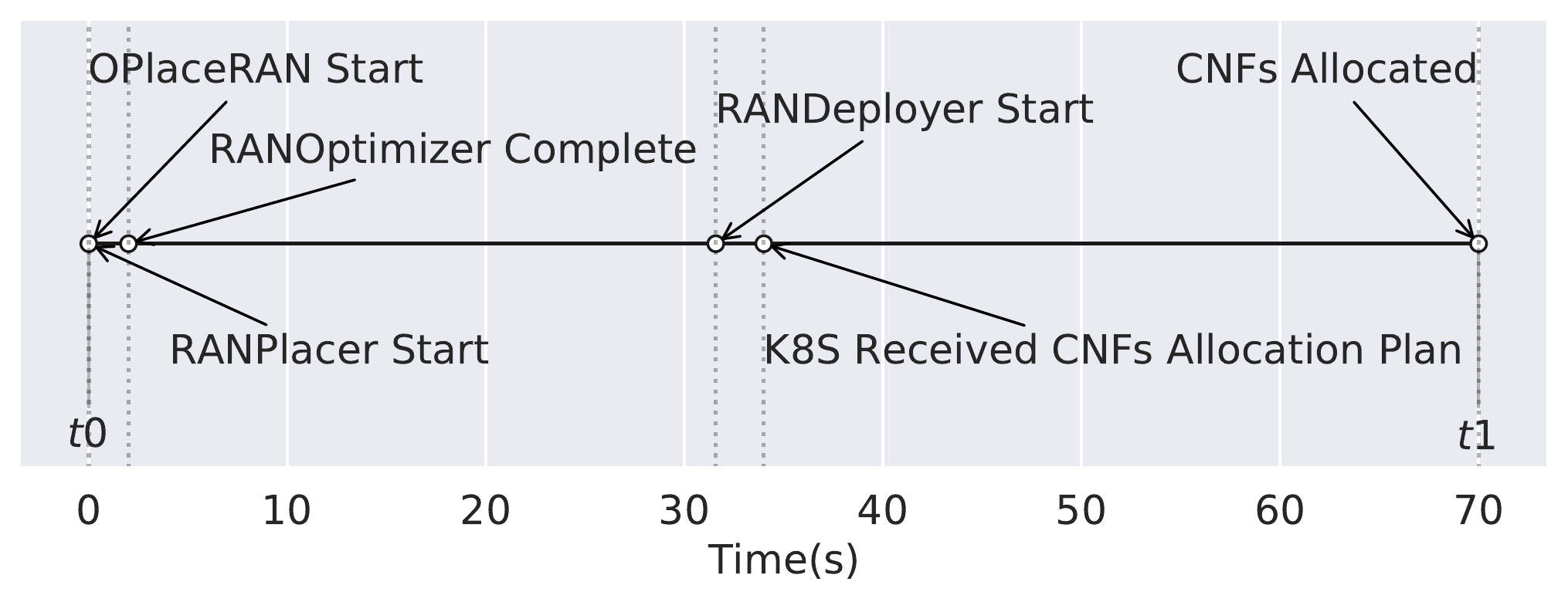}
    \end{center}
\end{figure}

According to the placement solution results, the analysis of cluster computing behavior shows the relationship between the main events during the deployment of OPlaceRAN based on the use of CRs of the cluster. In this analysis, we only evaluate the PlaceRAN placement solution once the focus is on the behavior of OPlaceRAN and CRs. The events for a full deployment are highlighted in seven moments of time ($t0$ up to $t7$) as shown in Table \ref{tab:operator_moments}. Initially, the time consumed by the entire OPlaceRAN, from the collection of the external inputs (sent by Network Operator) up to the allocation of CNFs, is presented between $t0$ and $t1$s shown in Fig. \ref{fig:operator-times}. The time between $t0$ and $t1$ last approximately 70s being divided into six main events. The first event occurs with OPlaceRAN Start, beginning with the Network Operator and immediately RANPlacer Start to manage the process. After that, the placement solution begins processing the inputs Network topology, vRUs locations, and node resources. The RANOptimizer Complete processing time takes approximately 1.2s after starting the placement due to the low complexity of the topology composed by three nodes. Following that, RANDeployer processes the placement solution result and sent for \textit{RANDeployer Start}. These steps take around 31.5s. In the next step, RANDeployer sends the plan to the K8S with the configuration to be applied. Finally, K8S allocates the CNFs into the Pods (vCUs, vDUs, and vRUs) images to run on K8S nodes. When K8S Received CNFs allocation plan occurs, nodes download the OAI images and  CNFs Pods are started (34s), the OAI is loaded by the container layer and, finally, all \textit{CNFs become allocated} (70s).


\begin{table}[h!]
    \centering
    \caption{Key Operator's activities in function of time.}
    \label{tab:operator_moments}
        \begin{tabular}{|c|c|}
            \hline
            \textbf{Time} & \textbf{Description}                                    \\ \hline
            \rowcolor[HTML]{cdd5e4} $t0$     & Started OPlaceRAN                                      \\
            $t1$     & Ended OPlaceRAN, CNFs allocated                        \\
            \rowcolor[HTML]{cdd5e4}  $t2$     & Configured OAI network parameters                     \\
            $t3$     & OAI layer is loaded                                   \\
            \rowcolor[HTML]{cdd5e4} $t4$     & OAI functional and connected to Core Network                  \\
            $t5$     & Configured network tunnel between vRU and Core Network \\
            \rowcolor[HTML]{cdd5e4} $t6$     & Ping from simulated UE to Core Network started                \\
            $t7$     & Ping from simulated UE to Core Network terminated             \\ \hline
        \end{tabular}
\end{table}

\begin{figure}[h!]
    \begin{center}
        \caption{Cluster Computing Resources usage behavior.}
        \label{fig:CNFs_CR}
        \includegraphics[width=.5\textwidth]{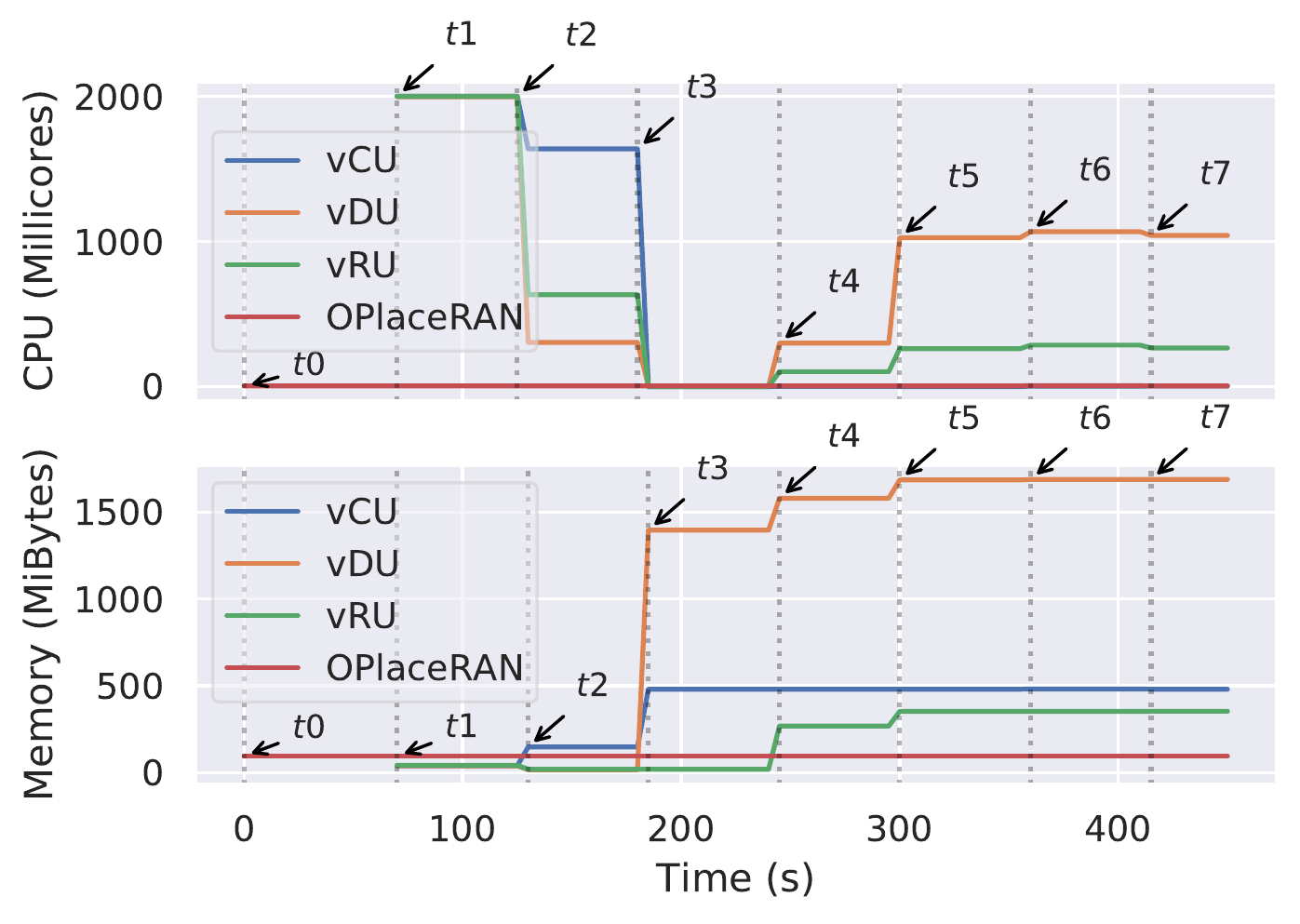}
    \end{center}
\end{figure}

After the OPlaceRAN tasks are performed, it is necessary to configure the radio units through the parameters provided by the OAI tool for each RAN function (i.e., vCU, vDU, vRU) after all Pods are allocated and before starting the RAN CNFs. This configuration happens between $t1$ and $t2$ using the maximum CPU, near to 2000 CPU Millicores (i.e., two vCPUs), shown at the top of Fig. \ref{fig:CNFs_CR}. After radio units are configured, the load of OAI layer is finished in $t3$. Next, OAI has already been configured and loaded, OAI processing is started. In this sense, the memory consumption in the K8S cluster increases significantly between $t3$ and $t4$, shown at the bottom of Fig. \ref{fig:CNFs_CR}. Following the $t5$ time represents the establishment of the tunnel between vRU and the Core Network. In the $t6$ event, a network tunnel between the simulated UE and Core Network is established, and data traffic from the simulated UE to Core Network using the Ping application starts with default configurations for the 60s. We can observe a slight increase of CPU at the top of Fig. \ref{fig:CNFs_CR} between $t6$ and $t7$. Finally, it is fundamental to highlight that the CPU and Memory consumption of OPlaceRAN is insignificant. We can see in Fig. \ref{fig:CNFs_CR} that the OPlaceRAN overhead hardly varies as a function of times and tasks running on the K8S cluster. In general, CPU and memory of vDU and vRU stabilize from $t5$ with the constant data packets generation, differently of vCU that did not vary after $t3$ because it only works as a traffic tunnel support for data packets after OAI is configured. One of the reasons for this behavior is that the OAI version works with emulated UE (without PHY protocol), and these UEs are fixed, reducing the network's signaling. This behavior was not considered as a trouble because of the paper focus.



\begin{figure}[h!]
    \begin{center}
        \caption{Nodes resources average behavior.}
        \label{fig:nodes_resources}
        \includegraphics[width=.5\textwidth]{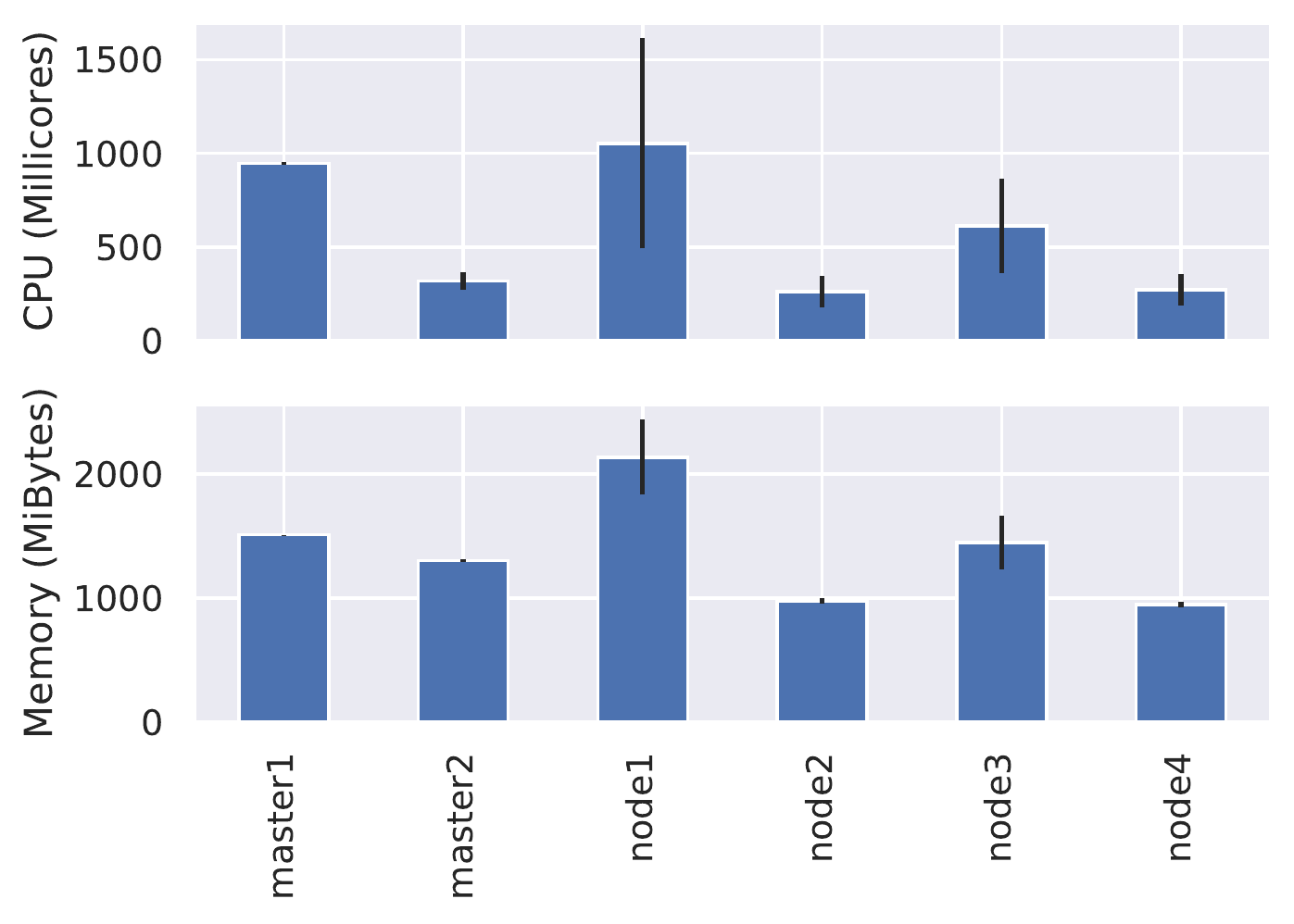}
    \end{center}
\end{figure}

We must consider Table \ref{tab:computing_resouces} to understand the CR consumption in each node of the K8S cluster. Firstly, we show the amount of CPU and memory that is guaranteed to be available to the CNFs placement, i.e., the input values to the PlaceRAN placement solution. The CRs diversity is necessary to produce placement solutions that have different behavior. In this case, the objective was to simulate an environment with scarce resources, with the placement solutions having RAN units behavior so diverse. Node 1 running in the network provided in Fig. \ref{fig:CNFs_CR} presents the highest average consumption of CRs during the experiment execution, 1053 CPU (Millicores) and 2135 MiBytes of Memory, as shown in Fig. \ref{fig:nodes_resources}. It makes sense because this node has higher CRs (5000 CPU Millicores and 2000 MiBytes of free memory), and all tested placement solutions prioritize to allocate the CNFs in Node 1. Both Nodes 2 and 4 allocated just one vRU because they have lower free CRs (1000 CPU Millicores and 200 MiBytes of Memory). Moreover, PlaceRAN allocated just one vRU because these CNFs use lower CRs than others, as seen in Fig. \ref{fig:operator-times}. Finally, Node 3 has an intermediate capacity of free CRs (2000 CPU Millicores and 1200 MiBytes of Memory). Furthermore, this node has a medium average consumption of CRs 613 CPU (Millicores) and 1448 MiBytes of Memory, but it has the highest allocation of proportional usage of CRs. This behavior is a consequence of the placement solution PlaceRAN because the Multi-CU vRAN allocates a lower number of CNFs in this Node 3. A relevant observation is the consumption of the Master nodes. Given the proof-of-concept scale, the four vRANs chains present in the cluster did not generate a high workload for the K8S control nodes. Master 1 has higher CPU and Memory usage because it is the primary K8S control plane.

\subsection{Discussion}
\label{discussion}



Initially, we highlight OPlaceRAN within its conceptual design, including the prototype description, and present the development of vNG-RAN. Four pillars are crucial for the solution to achieve great potential: (\textit{i}) alignment and positioning with the up-to-date architecture and frameworks of vNG-RAN, (\textit{ii}) agnostic placement of CNFs approach, (\textit{iii}) real prototype update with main tools and emulators, and crosshaul topology analysis. Furthermore, the first item opens opportunities for the integration of OPlaceRAN to any other solutions also aligned with the NFV architecture and SMO framework, whether this is within the scope of vNG-RAN itself or with the CNs’ crosshaul transport networks.



Concerning the possibility of operating in an agnostic way for placement approaches, OPlaceRAN stands as a unique solution. Such an operation provides an opportunity to analyze different placement solutions, as performed in Subsection \ref{subsec:results}, which compares two different strategies for the vNG-RAN placement. Furthermore, the possibility of exploring different placement solutions brings near-real operational network results. This approach differs from traditional mathematical placement solutions as they do not evaluate realistic NFVI infrastructures and containerized RAN applications. In this sense, i.e., the study of the NFVI cluster and its respective nodes, Subsection \ref{subsec:results} makes it possible to add an assessment of the impact on CRs based on their hardware, operating system, functions, and consumption characteristics of native tools (e.g., K8S Master control and CNI Calico), and the deployment of the CNF allocation concurrently. Therefore, to bring a real impact to the performance of NFVI and RAN CNFs, in some cases, demands reassessments in the mathematical exact and heuristic approaches of placement solutions.



Finally, the interconnection of vNG-RAN through crosshaul networks, currently one of the biggest challenges of disaggregated RAN, can be performed by OPlaceRAN. Due to the size of the topology investigated in the proof-of-concept, the chaining of CNFs vCUs, vDUs, and vRUs present very similar latency values. Overall near 18ms on average for the three scenarios presented by the placement in the end-to-end communication approaches between UE and CN (analysis performed within the mobile network, not evaluated in external connections). Two relevant aspects of the latency’s similarities are the number of hops (low processing latency) and the physical distance between nodes (all nodes in the same environment). Furthermore, since during the tests occurred various latency peaks, to understand this, we have to consider certain aspects: (\textit{i}) we calibrated and left some slack the CRs available for the Core Network and RAN Pods, and (\textit{ii}) we customized the OAI version as container-based with two splits disabling the PHY protocol. Finally, we have not investigated the bandwidth parameter in the evaluation because the maximum obtainable value is not relevant to generate an impact on the network (around 150 Mbps at maximum capacity, as shown in Table \ref{tab:profileslte}). However, OPlaceRAN supports bandwidth configuration.

\section{Related Work}
\label{sec:rw}

Recently, RAN faced an intense process of \textit {softwarization} and virtualization, leading to a broad transformation. In this scenario, the orchestration process has great relevance for the development of vNG-RAN. However, the research about virtualized RAN network functions orchestration has generated a divergent view between several works. Mainly concerning supporting the characteristics and requirements of vNG-RAN and the subsequent placement solutions strategies in the experimental orchestrators. In this context, the experimental orchestration studies are classified into investigations by open-project organizations and academia. For open-source projects, a strong alignment with the NFV architecture and the RAN disaggregation is observed. The projects Open Network Automation Platform (ONAP) \cite{onap}, Open Source MANO (OSM) \cite{osm}, Mosaic 5G \cite{nikaein2018mosaic5g} stand out in this scenario, led by the organizations Linux Foundation, European Telecommunications Standards Institute (ETSI), and EURECOM, respectively. For example, the project under development by ONAP is natively oriented by O-RAN initiative. Nevertheless, open-source projects still lack concepts aware of placement and crosshaul networks.

For academic purposes, we developed Table \ref{tab:rw} to guide the analysis of works, showing the state-of-the-art orchestration, placement solutions, NFV architecture, CNF support, and RAN disaggregation. The orchestration characteristic is the most relevant when comparing the related work since there are investigations with different goals. In some studies, orchestration focuses on the fronthaul performance \cite{makris2019virtualized,kondepu2018orchestrating,saha2018novel}, while other works prove concepts related to the framework's functioning without an orchestration strategy \cite{kondepu2018orchestrating,seung2018virtualization,rodriguez2020cloud}. However, Dalla-Costa et al. \cite{dalla2020orchestra} and Matoussi et al. \cite{matoussi20205g} provide consistent orchestration goals. The first work introduces a load balancing algorithm with an analysis of the performance of the fronthaul network and the computational resources. The authors use a synthetic C-RAN, with the behavior results closely regular and constant. The second work developed a particular heuristic approach for the placement of radio functions to reduce the consumption of computational resources.
Based on the works presented, OPlaceRAN is the first work in the literature that supports agnostic placement solutions, enabling different placement strategies, empowering the planning of vNG-RAN. Moreover, OPlaceRAN deals in an integrated way with CRs strategies and topologies features for crosshaul networks.

\begin{table*}[!ht]
    \centering
    \caption{Related Work - Academic Placement Orchestrators.}
    \label{tab:rw}
    \resizebox{\textwidth}{!}{
        \begin{tabular}{|c|c|c|c|c|c|}
             \hline
            \textbf{Works} &
              \textbf{\begin{tabular}[c]{@{}c@{}}Orchestration\\ Goals\end{tabular}} &
              \textbf{\begin{tabular}[c]{@{}c@{}}Placement\\ Solution \end{tabular}} &
              \textbf{\begin{tabular}[c]{@{}c@{}}NFV\\ Architecture\end{tabular}} &
              \textbf{\begin{tabular}[c]{@{}c@{}}CNF\\ Support\end{tabular}} &
              \textbf{\begin{tabular}[c]{@{}c@{}}RAN\\ Disaggregation\end{tabular}} \\
             \hline
             \rowcolor[HTML]{cdd5e4}
              \cite{dzogovic2019connecting} &
              Native K8S tool orchestration &
              \xmark &
              \xmark &
              \cmark &
              Not specified \\ 
              \cite{kondepu2018orchestrating} &
              High availabilty fronthaul orchestration &
              \xmark &
              \cmark &
              \cmark &
              C-RAN \\ 
              \rowcolor[HTML]{cdd5e4}
              \cite{saha2018novel} &
              Fronthaul driven RAN &
              \xmark &
               \xmark &
              \xmark &
              C-RAN \\ 
              \cite{seung2018virtualization} &
              Manual orchestration of VNF and PNF &
              \xmark &
              \cmark &
              \xmark &
              NG-RAN 2 units \\ 
              \rowcolor[HTML]{cdd5e4}
              \cite{rodriguez2020cloud} &
              Demo orchestration under ONAP platform &
              \xmark &
              \cmark &
              \cmark &
              NG-RAN 3 units \\ 
              \cite{makris2019virtualized} &
              Wireless fronthaul performance driven orchestration &
              \xmark &
              \cmark &
               \xmark &
              NG-RAN 2 units \\ 
              \rowcolor[HTML]{cdd5e4}
              \cite{dalla2020orchestra} &
              Resources load balancing orchestration for C-RAN &
              Specific &
              \cmark &
              \cmark &
              C-RAN \\ 
              \cite{matoussi20205g} &
              Specific placement orchestrator for C-RAN  &
              Specific &
              \xmark &
              \cmark &
              C-RAN \\ 
             \rowcolor[HTML]{cdd5e4}
              OPlaceRAN &
              Agnostic placement orchestrator for vNG-RAN &
              Agnostic &
              \cmark &
              \cmark &
              NG-RAN 3 units \\ 
              \hline
        \end{tabular}
    }
\end{table*}

Two studies about virtualization are aligned with the NFV architecture and support for container technology. The NFV architecture is an essential guide to the virtualization process because its reference model integrates several parts of fifth-generation of mobile networks, e.g., CN and RAN integration  \cite{gsmamigration}. Moreover, container virtualization support is the leading solution for the granularity of virtualized radio functions \cite{Gavrilovska2020}. In this sense, Rodriguez et al. \cite{rodriguez2020cloud} present a demonstration in line with the NFV architecture and meet container virtualization under the K8S tool. However, the authors only provide the RAN disaggregation for NG-RAN with three radio units guided by O-RAN concepts, without taking into account the crosshaul networks, and do not consider any placement solutions. Therefore, despite all the ongoing development on virtualized architectures for the fifth-generation mobile networks, OPlaceRAN fills an open research gap regarding an orchestrator with vNG-RAN placement support aligned with NFV architecture, directives of the O-RAN projects, and evolution from RAN to vNG-RAN. Furthermore, OPlaceRAN is aligned with up-to-date virtualization tools and technologies based on CNFs, including functional RAN splitting requirements, awareness of the crosshaul topology and computational resources, and providing opportunities for the integration with other platforms further than the mobile networks.


\section{Final Remarks}
\label{sec:conc}


In this work, we have proposed and validated the OPlaceRAN solution for orchestration of placement optimization solutions of virtualized disaggregated radio functions guided by the standards defined for vNG-RAN and aware of the requirements crosshaul networks and computational resources. Moreover, the OPlaceRAN concept has native support for generic and agnostic placement solutions approaches, providing opportunities for different solutions strategies to leverage the RAN planning evaluations. Furthermore, the concept aligned with the NFV architecture and set with the O-RAN framework brings an up-to-date position of the work in the RAN research. Finally, a prototype based on the cloud-native, CNFs concepts, and K8S tool are developed to validate the conceptual orchestrator.



OPlaceRAN was evaluated in an experimental scenario based on the deployment of two different placement solutions in the proof-of-concept. In this sense, we proved an effective and agnostic optimization model in a testbed experiment, emulating a mobile network with open-source tools. The conceptual solution and experimental results showed the effectiveness of OPlaceRAN in the orchestration of control and deployment of placement solutions strategies for vNG-RAN. The entire orchestration steps, cluster/nodes resources, and crosshaul requirements are controlled by the OPlaceRAN orchestrator with a low computational overhead. Finally, we can conclude that OPlaceRAN contributes to the vNG-RAN transformation and opens opportunities to integration with other fifth-generation of mobile network platforms.



For future work, the OPlaceRAN opens several opportunities. First, we envision adding to the orchestrator an integration of flow forwarding controller by the Software-defined Networking (SDN) to leverage and automatize the crosshaul network’s capacity and evaluate OPlaceRAN in a topology with more nodes and complexity. The other relevant goal is to connect the core network standalone standard and integrate it with RAN placement strategies. Moreover, there is the need for introducing the RAN slicing concepts aligned with O-RAN.


\section*{Acknowledgment}

This work was conducted with partial financial support from the National Council for Scientific and Technological Development (CNPq) under grant number 130555/2019-3 and from the Coordination for the Improvement of Higher Education Personnel (CAPES) - Finance Code 001, Brazil.


\bibliographystyle{IEEEtran}
\bibliography{biblio}

\begin{IEEEbiography}[{\includegraphics[width=1in,height=1.1in,clip]{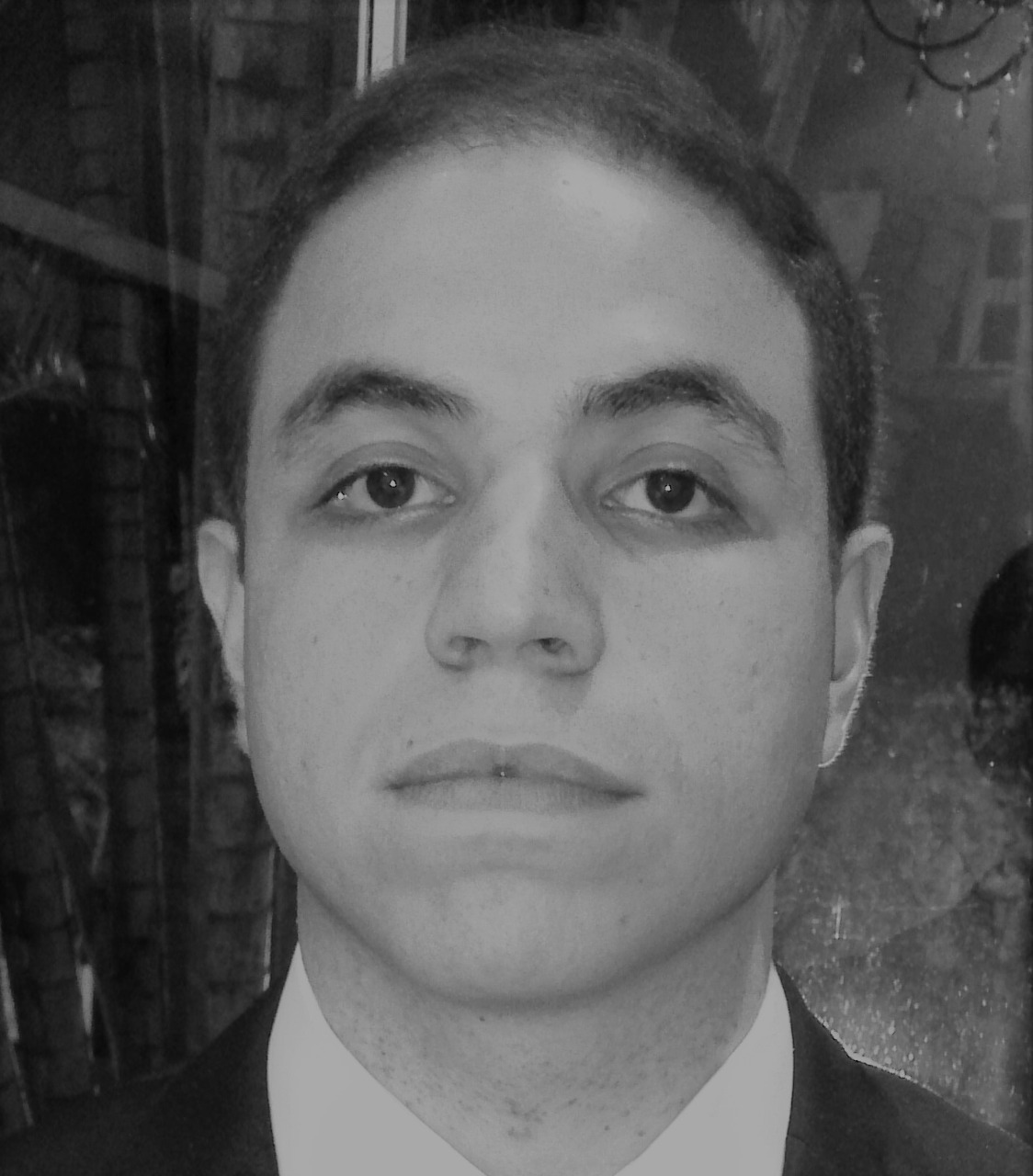}}]{Fernando Zanferrari Morais} is MSc degree at University of Vale do Rio dos Sinos (UNISINOS), Brazil. Fernando received the bachelor's degree from Centro Universitário Lasalle in 2009. He served as a Telecommunication Engineer at Vivo Operator, a Telefonica group, Brazil. His research interests include telecommunication mobile networks, software-defined networking, virtualization and cloud computing. \end{IEEEbiography}

\vspace{-1.5cm}

\begin{IEEEbiography}[{\includegraphics[width=1in,height=1.1in,clip]{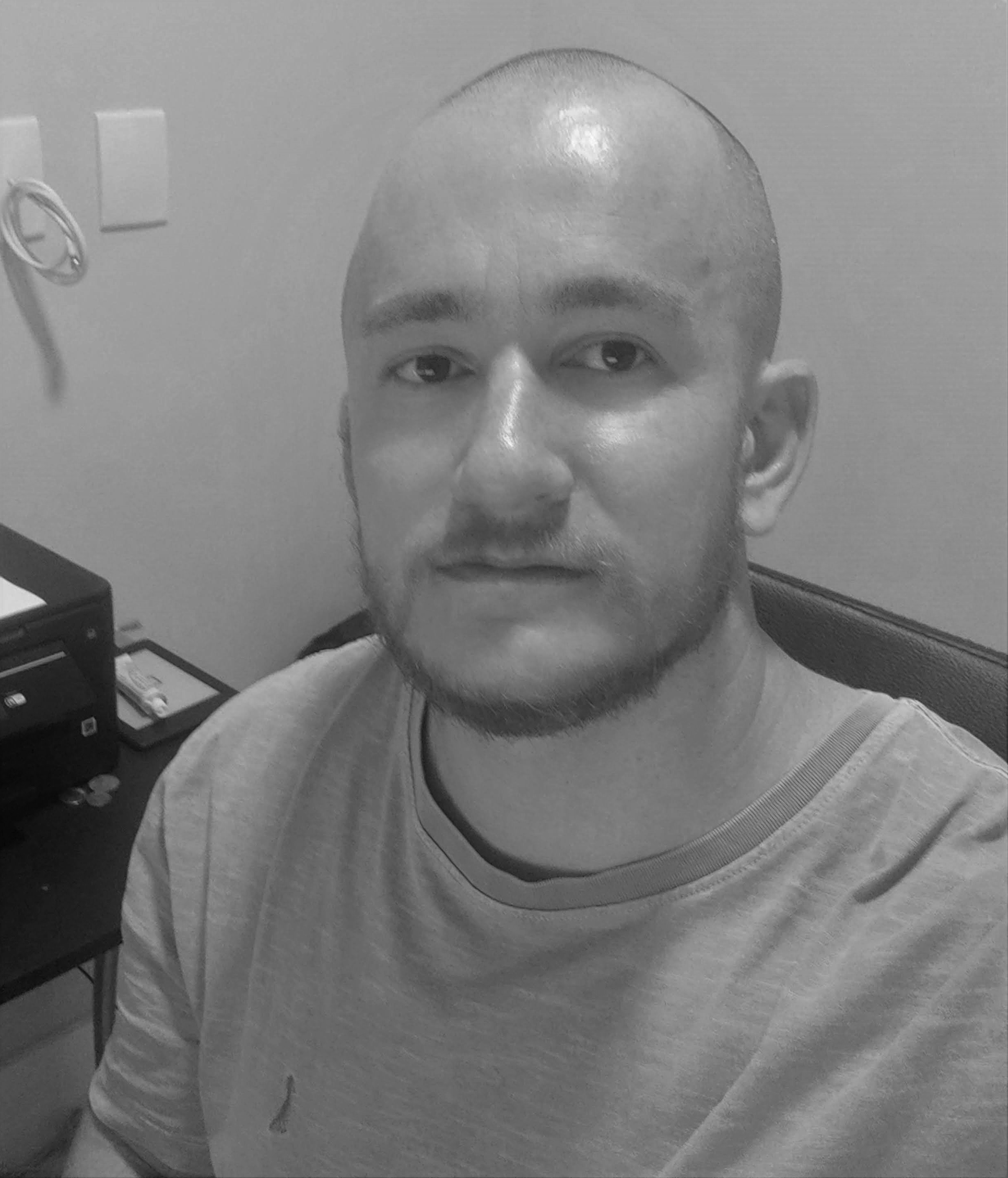}}]{Gustavo Zanatta Bruno} is a Ph.D. student in Computer Science at the University of Vale do Rio dos Sinos (UNISINOS) and MSc in Computer Science at Universidade Federal Fluminense (UFF), both in Brazil. He served as an Information Technology Infrastructure Analyst at MTI, a public company responsible for IT services in Mato Grosso, Brazil. His research interests include infrastructure automation, mobile telecommunication networks, and cloud computing.\end{IEEEbiography}

\vspace{-1.5cm}

\begin{IEEEbiography}[{\includegraphics[width=1in,height=1.1in,clip]{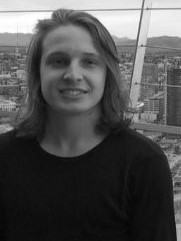}}]{Julio Renner} is a Software Developer at Routific, Canada. Graduated in Computer Science at the University of Vale do Rio dos Sinos (UNISINOS), Brazil. His research interests include software containerization and orchestration, DevOps and cloud computing. \end{IEEEbiography}

\vspace{-1.5cm}

\begin{IEEEbiography}[{\includegraphics[width=1in,height=1.1in,clip]{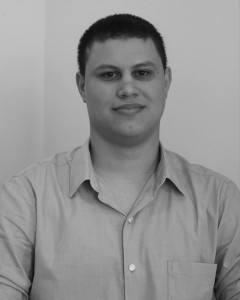}}]{Gabriel Matheus F. de Almeida} has been a Computer Science Researcher and a member with the Laboratory Computer Networks and Distributed Systems, Federal University of Goiás (UFG), Brazil, since 2019. His research interests include  wireless networks, software-defined networks, virtualization, resource allocation and performance evaluation. \end{IEEEbiography}

\vspace{-1.5cm}
 
\begin{IEEEbiography}[{\includegraphics[width=1in,height=1.1in,clip]{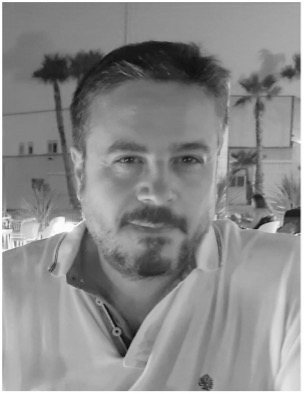}}]{Luis M. Contreras} earned a Telecom Engineer (M.Sc.) degree at the Universidad Politécnica of Madrid (1997), and holds an M.Sc. on Telematics from the Universidad Carlos III of Madrid (2010). Since August 2011 he is part of Telefónica I+D / Telefónica CTIO unit, working on 5G, SDN, virtualization, and transport networks. He is part-time lecturer at the Universidad Carlos III of Madrid, where is also pursuing a Ph.D. (expected for 2021).  \end{IEEEbiography}

\vspace{-1.5cm}

\begin{IEEEbiography}[{\includegraphics[width=1in,height=1.1in,clip]{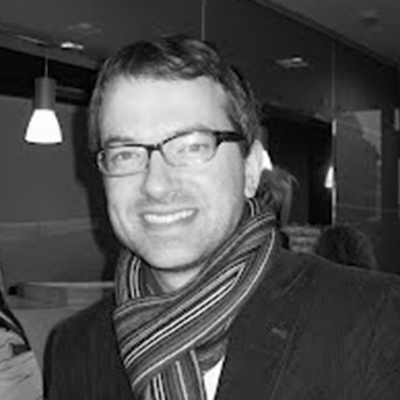}}]{Rodrigo da Rosa Righi} is professor and researcher at University of Vale do Rio dos Sinos (UNISINOS), Brazil. Rodrigo concluded his post-doctoral studies at KAIST - Korean Advanced Institute of Science and Technology, South Korea. He obtained his MS and PhD degrees in Computer Science from the Federal University of Rio Grande do Sul, Brazil, in 2005 and 2009, respectively. He is a member of both IEEE Computer Society (senior member) and ACM. \end{IEEEbiography}

\vspace{-1.5cm}

\begin{IEEEbiography}[{\includegraphics[width=1in,height=1.1in,clip]{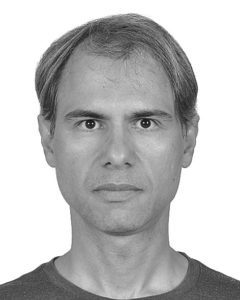}}]{Kleber Vieira Cardoso} is an associate professor at the Institute of Informatics – Universidade Federal de Goiás (UFG), where he has been a professor and researcher since 2009. He holds a degree in Computer Science from UFG (1997), has MSc (2002) and PhD (2009) in Electrical Engineering from COPPE – Universidade Federal do Rio de Janeiro. In 2015, he spent his sabbatical at Virginia Tech (in the USA) and, in 2020, at Inria Saclay Research Centre (in France). \end{IEEEbiography}

\vspace{-1.5cm}

\begin{IEEEbiography}[{\includegraphics[width=1in,height=1.1in,clip]{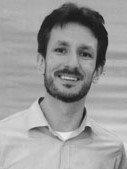}}]{Cristiano Bonato Both} is a professor of the Applied Computing Graduate Program at the University of Vale do Rio dos Sinos (UNISINOS), Brazil. He coordinators research projects funded by H2020 EU-Brazil, CNPq, FAPERGS, and RNP. His research focuses on wireless networks, next-generation networks, softwarization and virtualization technologies for telecommunication networks. \end{IEEEbiography}

\end{document}